\documentclass[11pt]{article}
\usepackage[margin=1in]{geometry}
\usepackage[backend=bibtex,firstinits=true,style=numeric-comp,date=year,sorting=none]{biblatex}
\renewbibmacro{in:}{}
\AtEveryBibitem{\clearlist{language}}
\usepackage{textgreek}

\usepackage{graphicx}
\usepackage{float}
\PassOptionsToPackage{hyphens}{url}\usepackage[colorlinks=false]{hyperref}
\usepackage{bookmark}
\usepackage[all]{hypcap}
\usepackage[version=4]{mhchem}
\usepackage[capitalise]{cleveref}

\usepackage{wrapfig}
\usepackage{sidecap}
\usepackage{subcaption}
\usepackage[labelfont={bf,footnotesize}, textfont=footnotesize, skip=8pt plus 2pt minus 2pt]{caption}

\usepackage{mathtools} 
\usepackage{amsmath}
\usepackage{amssymb}
\usepackage{amsfonts}
\usepackage{xspace}
\usepackage{changepage}
\usepackage{breqn}

\usepackage{cleveref}
\usepackage{textcomp}

\usepackage{comment}

\usepackage{array}
\usepackage{booktabs}  
\usepackage{multicol}
\usepackage{multirow}

\usepackage{adjustbox}

\usepackage{siunitx}
\DeclareSIUnit{\molecule}{molecule}
\DeclareSIUnit{\molecules}{molecule}
\DeclareSIUnit\molar{\mole\per\cubic\deci\metre}
\DeclareSIUnit\Molar{\textsc{M}}
\usepackage[shortlabels]{enumitem}
\setlist{nosep} 

\usepackage{xcolor}

\usepackage[color=green!20]{todonotes}

\newcommand{\thinsim}{{\raise.17ex\hbox{\(\scriptstyle\mathtt{\sim}\)}}}
\newcommand{\PR}[1]{{\color{purple}#1}}
\usepackage[flushleft]{threeparttable}

\usepackage[acronym, nonumberlist, nopostdot, nogroupskip, numberedsection=false]{glossaries}
\setacronymstyle{long-short}
\makeglossaries{}

\begin{document}
\begin{centering}
    \textbf{\Large Interplay between cortical adhesion and membrane bending regulates microparticle formation}\\[3mm]
    \textbf{A. Mahapatra$^{1, 2}$, S. A. Malingen$^{1,3}$, and P. Rangamani$^{1,**}$}\\[1mm]
    $^{1}$ Department of Mechanical and Aerospace Engineering,\\
    University of California San Diego, La Jolla CA 92093.\\
    $^{2}$ Present address: Senior Research and Development Engineer, Gamma Technologies, LLC, Westmont, IL  \\
    $^{3}$ Present address: Department of Bioengineering, University of Washington, Seattle \\
    $^{**}$To whom correspondence must be addressed: prangamani@ucsd.edu\\
\end{centering}




\section*{Abstract}
The formation of microparticles requires the bending of the plasma membrane away from the cytosol.
The capcity of the cell membrane to form a microparticle, and the rate of membrane deformation, are controlled by multiple factors, including loss of lipid asymmetry (primarily the exposure of phosphatidylserine on the outer leaflet), detachment of the membrane from the cortical cytoskeleton, and bleb expansion due to pressure.
In this work, we develop a biophysical model that accounts for the interaction between these different factors. 
Our findings reveal that the linkage between the membrane and cortex is a key determinant of outward budding. 
Ultimately, the rates and mechanical parameters regulating cortical linkage interact with the kinetics of phosphatidylserine flipping, laying out a mechanical phase space for regulating the outward budding of the plasma membrane. 

\section*{Significance Statement}
Cells release membrane-enclosed vesicles that can serve a role in long-range signalling. These extracellular vesicles are released in response to stress, inflammation,  injury and chemoresistance, but also in homeostatic regulation. One critical role of EVs is in the coagulation cascade.
A particular class of vesicles called ectosomes or microparticles are released by the outward budding of the plasma membrane, a process which requires detachment of the membrane from the cortex, exposure of negatively charged, curvature-inducing lipids such as phosphatidylserine from the inner leaflet to the outer leaflet, and pressure-driven bleb expansion. 
Here, we used membrane mechanics coupled with the kinetics of adhesive linker binding-unbinding to investigate how these different factors interact together. 
Using our model, we predict how linker properties influence outward budding of the plasma membrane and identify conditions that can promote or inhibit membrane curvature generation. 
These findings provide insight into the fundamental mechanisms underlying microparticle formation, elucidating the basic biology of processes like the procoagulation cascade.
Further, these mechanistic insights may inspire techniques for inhibiting microparticles where they are harmful, such as in cases of chemoresistant drug efflux by tumor cells. 

\section*{Keywords}
Microparticles; Helfrich energy; Adhesion; Ezrin linkers; Phosphatidylserine-induced spontaneous curvature.

\section*{Introduction}

Many cells in the animal body release extracellular vesicles from their cell surface under normal and stress conditions \cite{Hargett2013-ut,Brett2019-an,Segawa2011-va}.
Circulating microparticles (MPs) are critical for cell-cell communication and are reflective of a state of inflammation in the body \cite{Puddu2010-bo}.
Conditions that introduce cell stress such as inflammation, proapoptotic conditions, and environmental stressors such as  hyperbaric pressure in divers can result in increase in microparticle formation \cite{Brett2019-an}.
To avoid confusion, in this work, when we say cellular microparticles, we are referring to those membrane-encased vesicles that are released from the cellular plasma membrane, also referred to as ectosomes or microvesicles.
These microparticles are fragments of the plasma membrane released in response to an activating stimulus such as oxidative stress, cytokines, mechanical forces or in proapoptotic cells.
MPs range in size from about 50 to \SI{1000}{\nano\meter} \cite{Freyssinet2003-fo}. 
Their formation is distinct from other membrane fusion processes such as multivesicular bodies fusing with the plasma membrane and releasing their content into the extracellular space. 

\begin{figure*}[t!]
    \centering
    \includegraphics[width=\textwidth]{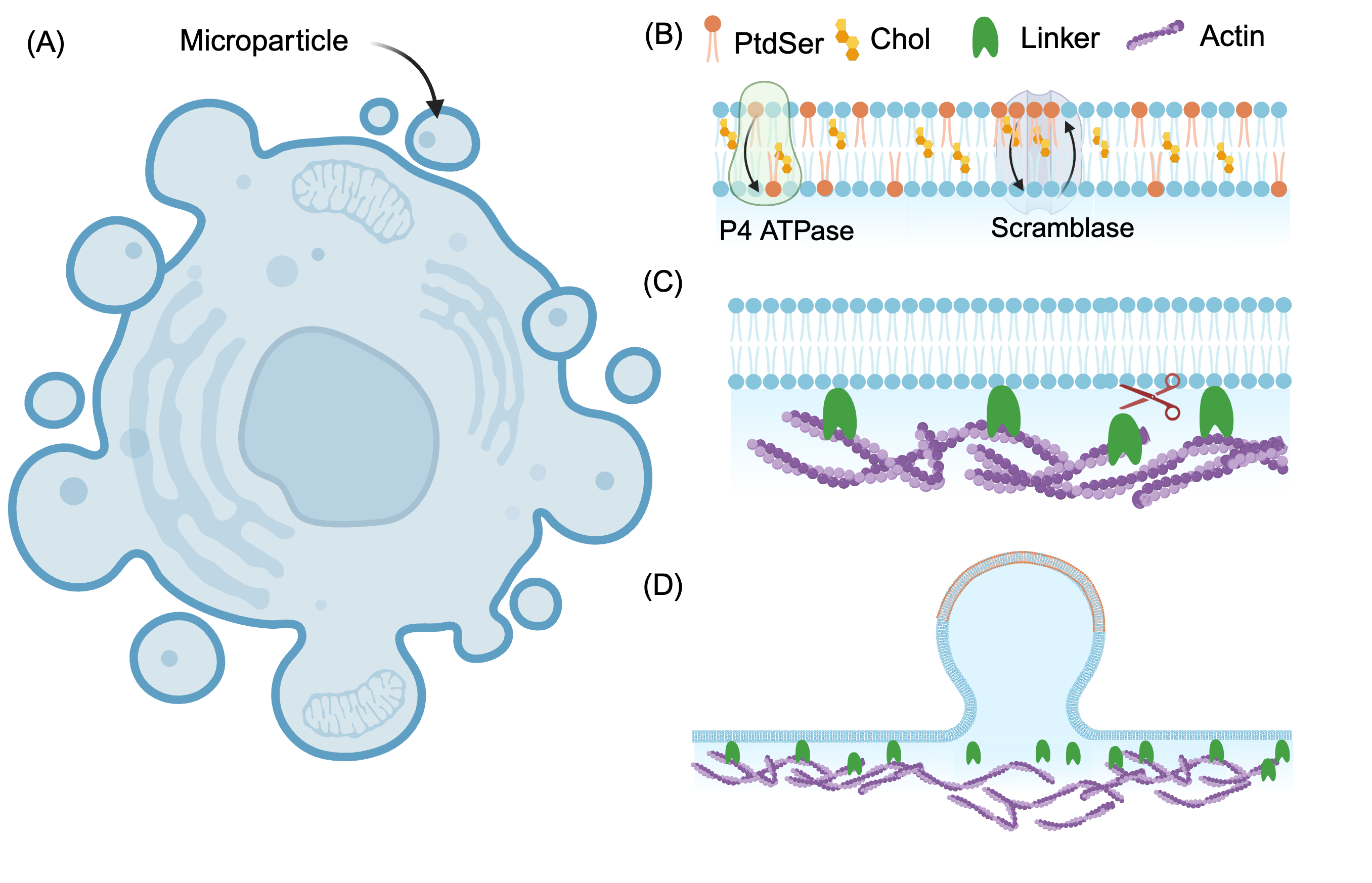}
    \caption{\textbf{Formation of microparticles on the cell surface is a biophysical process.} A) A schematic depicting microparticles being released from the plasma membrane of a cell. (B) A key initiator of microparticle formation is the exposure of phosphatidylserine on the plasma membrane, which is usually on the inner leaflet. P4ATPases are the flippases that flip the PS on the outer leaflet back to the inner leaflet and scramblases can promote lipid translocation between both leaflets. During microparticle formation, the scramblase TMEM16F translocates PS to the outer leaflet. (C) The cortical cytoskeleton is attached to the cell membrane by linkers, including those of the ezrin-radixin-moesin family. Detachment of the linkers from the membrane is a critical step for membrane deformation. (D) In this work, we model the formation of a microparticle by considering the combined role of the PS microdomain-induced curvature, detachment of the cortex by the linkers breaking, and pressure driven blebbing of the membrane. This figure was created using Biorender.com. }
    \label{fig:MicroparticleSketch}
\end{figure*}

Microparticles were first observed to form in red blood cells in the early 1900s \cite{Auer1933-rt,Kite1914-lk,Waitz1936-mt}.
In the 1970s, a series of studies showed that MPs from normal human red blood cells could be induced by elevation of intracellular calcium which can lead to alterations in the lipid profile of the erythrocyte membrane, and release of spectrin-free vesicles \cite{Allan1975-bp,Allan1976-ai,Lutz1977-iw}.
The formation of microparticles in red blood cells is also associated with aging and maturation of these cells \cite{Westerman2016-rb,Muller1994-br}.
Over the years, many different cell types including platelets and endothelial cells have been known to release MPs and MPs have been implicated in many different physiological and pathological conditions \cite{Ruhela2020-ks,Ruhela2021-jd,Distler2005-xq} (\Cref{fig:MicroparticleSketch}A). 

Many different experimental studies have characterized the different steps in MP formation and we briefly summarize them here grouped by either membrane effects or by cytoskeletal effects.
One of the hallmarks of MP formation is phospholipid asymmetry in the plasma membrane (\Cref{fig:MicroparticleSketch}B). 
Influx of local $\rm Ca^{2+}$ can induce MP formation by disrupting the transporters that, in a resting state, maintain the asymmetry of the leaflets \cite{Hugel2005-ri,Bevers2016-mw, Pasquet1996-cj}.
Phosphatidylserine, a negatively charged phospholipid, which is a known marker of apoptosis, has also been found to externalize from the inner to the outer leaflet of the membrane in a host of reversible processes unrelated to cell death \cite{Savitskaya2022-kl}. 
The timescale of PS exposure and its localization varies with the process, with a longer timescale ($\sim$ 1 hour) and global exposure during apoptosis and localized exposure driven by $\rm Ca^{2+}$ signalling on the order of 5-15 minutes \cite{Savitskaya2022-kl} in non-apoptotic events.
In fact, the exposure of PS on the outer leaflet is characteristic of MPs and the regions of the plasma membrane that eventually bud off into MPs are shown to be rich in PS on the outer leaflet as observed by Annexin V binding \cite{Hugel2005-ri, Freyssinet2003-fo}. 
Significantly, it has been demonstrated that cells can adjust the preferred curvature of their membrane using PS since the negatively charged head groups of this anionic phospholipid repel one another, generating a preferred curvature \cite{Hirama2017-lr}. 
The generation of lipid asymmetry on the bilayer with PS exposure on the outer leaflet has been extensively reviewed in the literature \cite{Clarke2020-re,Sakuragi2023-ys,Balasubramanian2007-gm,Bevers2016-mw}.

Another important feature in MP formation is the disruption of the links between the cortical cytoskeleton and the plasma membrane (\Cref{fig:MicroparticleSketch}C). 
Calcium signalling can also disrupt the integration of the membrane and cortex. 
$\rm Ca^{2+}$ activates calpain, which can disrupt linkages between the membrane and the cortex \cite{Weber2009-tw}.
Activated calpain can cleave the cortex, spurring cytoskeletal remodeling \cite{Randriamboavonjy2012-qt}. 
Loss of membrane-cytoskeleton integration can reduce the rigidity of the membrane \cite{Pasquet1996-cj}, in addition to reducing the adhesive force. 
Taken together, the delamination of the membrane from the cortex could make it locally more susceptible to hydrostatic pressure, which is a driving force for bleb formation. 
The formation of blebs in cell membranes is accompanied by the loss of membrane cytoskeleton adhesion \cite{Dai1999-fc}. 
Recent experiments have shown that for a cell protrusion to form, actin polymerization alone is not sufficient. 
In fact, breakage of the actin membrane links, specifically, ezrin, is a required condition for cell protrusion -- increases in actin-membrane attachment decreases protrusions while a decrease in actin-membrane attachment increases protrusions \cite{Welf2020-hx}.
Ezrin and related proteins are also involved in MP formation \cite{Perez-Cornejo2012-gy}, suggesting that loss of membrane-cytoskeleton attachement is a critical first step in MP formation.

Thus, from a biophysical standpoint, the formation of MPs can be characterized by three main driving forces: changes to membrane composition and accompanying changes to spontaneous curvature due to PS flipping, loss of membrane-cortex attachment by breaking the adhesive bonds, and pressure-driven bleb growth (\Cref{fig:MicroparticleSketch}D).
Each of these processes has been studied using mathematical models.
Formation of membrane curvature due to lipid-composition induced spontaneous curvature is done using the Helfrich model \cite{Hassinger2017-su,Helfrich1973-lr} and the specific role of PS has been recently incorporated in these models \cite{Hirama2017-lr}.
The role of pressure driven bleb growth has been studied using computational fluid dynamics approaches, but the role of membrane bending was not considered \cite{Strychalski2021-fo}.
Finally, a biophysical model of membrane-cortex linkage examined the relationship between force, linker stiffness, and membrane curvature in the regime of small deformations \cite{Alert2015-zq}.
In this work, we develop a combined model of these three factors and systematically investigate the role of linker adhesion, external loading such as pressure and force, and dynamics of the change of lipid composition in the formation of MPs. 
Specifically, we focus on the following questions: How does pressure or a point load interact with linker binding to lead to outward bending? What is role played by the kinetics of PS flipping?
Our model predicts that linker binding is a key determinant of MP formation. 
Strong linker binding can override other mechanisms of membrane deformation including PS-induced spontaneous curvature and pressure-driven blebbing. 
These findings play a critical role in integrating myriad experimental observations in the literature and providing a unifying mechanical framework for outward budding of the membrane.

\section*{Results}
We analyze the formation of MPs in response to different biophysical driving forces using our model.
Even though the individual elements of our model have been analyzed previously, we study on the coupled nonlinear system and focus on the formation of highly curved structures and the interplay between the kinetics of PS flipping and the kinetics of adhesive linkers.
To our knowledge, previous studies have not studied the combined elements in a nonlinear regime and our analysis provides new insights into how cortical linkage and PS flipping may interact to form MPs.

\subsection*{Qualitative description of the model}
In this work, the formation of microparticles is modeled as the outward bending of the membrane (away from the cytosol) in response to the accumulation of phosphatidylserine (PS) on the outer leaflet.
The biophysical model contains three main modules. \textbf{Module 1:} describes how the membrane bends in response to mechanical loading such as pressure or spontaneous curvature and is modeled using the Helfrich free energy; \textbf{Module 2:} kinetics of cortical linker binding and unbinding to capture the effects of the ERM family of proteins; \textbf{Module 3:} the kinetics of PS flipping onto the outer leaflet of the membrane due to the activity of the scramblases (see \Cref{fig:MicroparticleSketch}).
The three modules are coupled through the impact of linker adhesion on the membrane energy and the spontaneous curvature induced by PS flipping, ultimately leading to membrane bending (\Cref{fig:MicroparticleSketch}D). 

\paragraph{Module 1: Membrane bending energy}
We assume that the membrane is a two-dimensional material with negligible thickness. 
We also assume that the membrane is inextensible in the plane and elastic in out-of-plane bending with uniform bending rigidity \cite{Steigmann1999-es}.
The membrane responds to any lipid asymmetry, encoded in the spontaneous curvature, $C$, externally applied force, $f_{pull}$, or pressure difference across the membrane, $p$, by bending (see Supplementary material for normal and tangential stress balances). 
The inextensibility of the membrane is captured by the tension, $\lambda$ \cite{Rangamani2014-tq, Rangamani2013-us,Rangamani2022-bv, Steigmann1999-es}.
The Helfrich Hamiltonian \cite{Helfrich1973-lr} describes the free energy density ($W$) of the membrane deformation modified with the work done by the cortex force ($\mathbf{f}$) on the membrane. The modified free energy reads as
\begin{equation}
\label{eqn:free_energy}
W = \kappa [H - C(\theta^{\alpha})]^{2} + \bar{\kappa}K + \lambda -\mathbf{f} \cdot \mathbf{z},
    \end{equation}
where $H$ is the mean curvature of the membrane, $K$ is the Gaussian Curvature, $\kappa$ is the bending modulus, and $\bar{\kappa}$ is the Gaussian modulus and $\mathbf{z}$ is the vertical displacement of the membrane. 

\paragraph{Module 2: Adhesion and detachment of linkers}
In order for the membrane to bend outward to form microparticles, studies have shown that detachment of the lipid bilayer from the actin cortex is an important first step \cite{Welf2020-hx,Charras2006-gr}.
The attachment of the membrane to the actin cortex is through linker proteins, and we model these linkers similar to the model proposed in \cite{Alert2015-zq,Sens2020-oe}.
The kinetics of linker binding and unbinding are given by
\begin{equation}
    \frac{d\phi}{dt}=k_{on}[1-\phi]-k_{off}(z)\phi,
    \label{eq:phi_eq}
\end{equation}
where $\phi$ is the fraction of bound linkers, $k_{on}$ is the association rate, and $k_{off}$ is the dissociation rate of the linkers to the membrane.
The rate of dissociation, $k_{off}$, increases with the magnitude of the force exerted by the cortex through Bell's law \cite{Bell1978-qp,Alert2015-zq} and is given by $k_{off}=k_{off}^0e^{\frac{\hat{f} \delta}{k_B T}}.$
Here $\delta$ is the bond length of the linkers and $\hat{f}=k_{lin}z$, where $k_{lin}$ is the stiffness of the linker springs.
The membrane shape and linker kinetics are coupled through the vertical distance $z$.
We note that while some models have proposed a coupling between bound linker density and spontaneous curvature \cite{Lin2023-la,Lin2018-jm}, experiments showed that ezrin did not induce any membrane curvature \cite{Tsai2018-op}.
Therefore, we did not couple linker density to spontaneous curvature in our model.


\paragraph{Module 3: Phospatidylserine flipping and generation of spontaneous curvature}
PS is a negatively charged lipid and we model the asymmetry introduced by these lipids using a spontaneous curvature, denoted by $C$, that is localized to the region of lipid asymmetry \cite{Xu2013-ps}.
As in previous works \cite{Liu2009-cp,Xu2013-ps,Hirama2017-lr}, we assume that the spontaneous curvature on the membrane is linearly proportional to the density of PS.
The density of PS on the outer leaflet of the membrane is regulated by lipid translocation between leaflets due to the activity of scramblases such as TMEM16F \cite{Feng2019-zm, Falzone2022-bi, Bevers2016-mw, Watanabe2018-gt} and lipid flipping from the outer leaflet to the inner leaflet due the P4ATPases (\Cref{fig:MicroparticleSketch}B) \cite{Tadini-Buoninsegni2019-es, Sakuragi2023-ys, Xu2013-ps, Miyata2022-vx, Devaux2008-wf}. 
Furthermore, TMEM16F appears to induce membrane thinning, promoting a curvature-dependent feedback loop \cite{Falzone2022-bi} and P4ATPases also appear to function in a curvature dependent manner \cite{Devaux2008-wf}. 
In our model, we incorporated curvature driven feedback for the rate constants of the scramblase and the P4ATPase based on the literature \cite{Devaux2008-wf,Falzone2022-bi} using an exponential function similar to \cite{Liu2009-cp}.
The rate of change of PS density on the outer leaflet is modeled as 

\begin{equation}
\frac{d [PS]}{d t}=\underbrace{\frac{k_{scramblase}}{K_m+ [PS]}e^{\alpha H(s)}[PS]}_{\begin{array}{c} 
\text { scramblase activity } 
\end{array}}-\underbrace{k_{P4ATPase}[PS]e^{\beta H(s)}.}_{\begin{array}{c}
\text {P4ATPase activity }
\end{array}}
\label{eq:PS_curvature_feedback}
\end{equation}
The first term accounts for increase in PS on the outer leaflet due to the activity of the scramblase, where $k_{scramblase}$ is a second order rate constant with units of 1/(\textmu M s).
The second term accounts for a first order flipping of PS from the outer leaflet to the inner leaflet.
In both these terms, $H(s)$ is the mean curvature and $\alpha$ and $\beta$ are effectively modulating the kinetic parameters based on the local curvature. 
When both $\alpha$ and $\beta$ are identically zero, the curvature-mediated feedback is zero in the model.
While the exact rates of the scramblases and P4ATPases are not known, as we will show later, we obtain these rate constants by fitting to experimental data.

\paragraph{Model summary:}
The formation of MPs is determined by balancing the force exerted by the linkers that tether the membrane to the cortex and the opposing membrane deformation driven by intracellular pressure and spontaneous curvature. 
Time dependence enters the model through the rate equation for the linkers and the rate equation for PS kinetics.
Thus, while the membrane itself is at mechanical equilibrium, the kinetics of linker binding-unbinding and rate of PS flipping will determine the temporal dynamics of MP formation. 
We assume that PS density and spontaneous curvature are localized on the area of the MP bud region, and the magnitude of spontaneous curvature is assumed to be uniform across the bud region.
For simplicity, we considered an axisymmetric patch of the membrane for our numerical simulations (\Cref{fig:schematic_supp}).
All model parameters, detailed derivations, and numerical schemes are given in the supplementary material.

\subsection*{Linker detachment is required for membrane deformation in response to an instantaneous load}
\begin{figure}[!!h]
    \centering
    \includegraphics[width=0.8\textwidth]{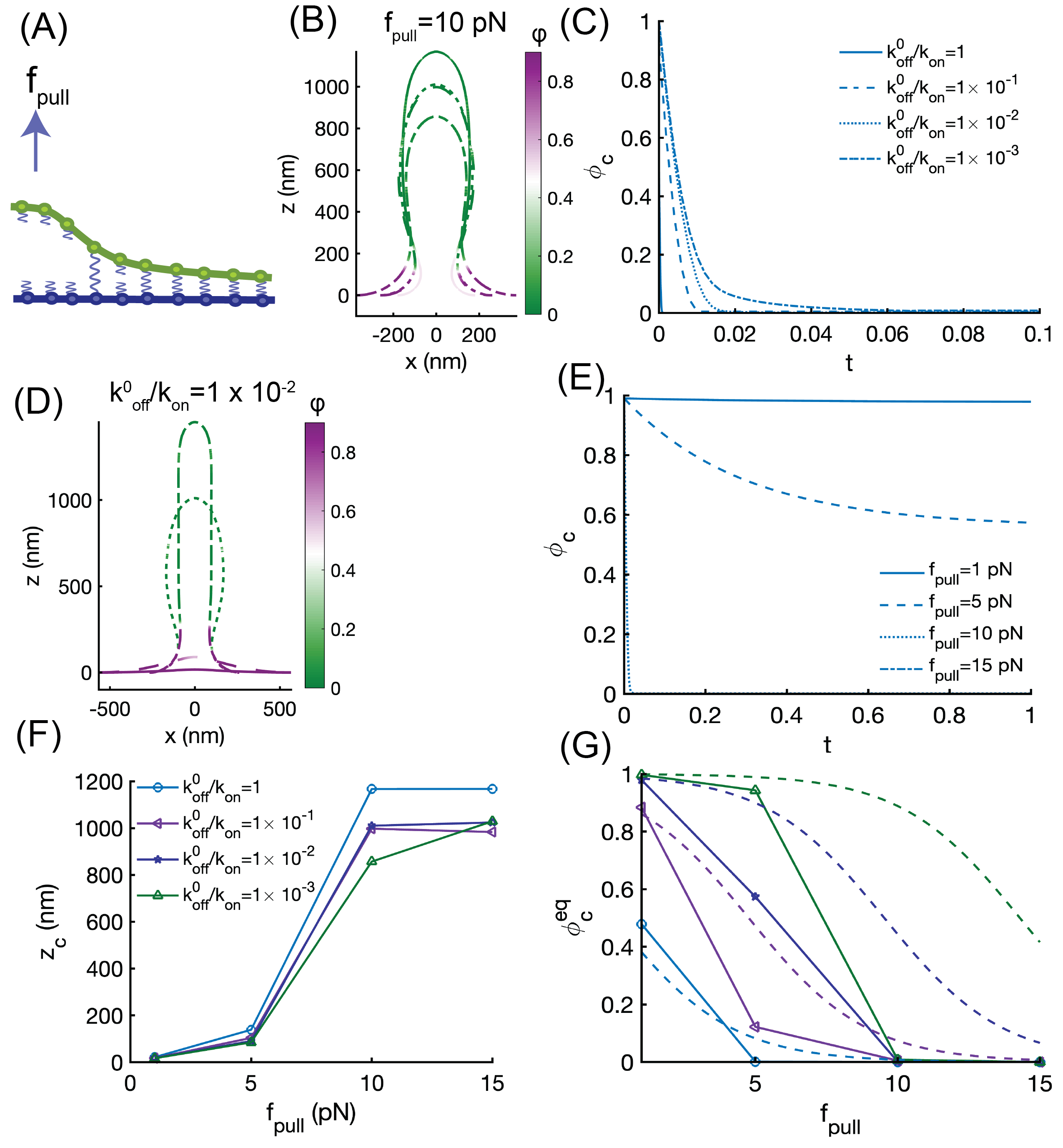}
    \caption{\textbf{Competition between linker binding and membrane bending determines tube formation.}
    (A) Schematic of linkers detaching in regions of applied pulling force in order for the membrane to detach. (B) Membrane shapes in response to $f_{pull} = 10$ pN for four different values of $\zeta=k_{off}^0/k_{on}$. The values of $\zeta$ are shown in panel (C).
    (C) Fraction of bound linkers in time at the center of the membrane for four different values of $\zeta=k_{off}/k_{on}$ and $f_{pull} = 10$ pN.
    The kinetics of $\phi$ are almost instantaneous for $\zeta=1$.
    (D) Membrane shapes in response to $k_{off}^0/k_{on} = 0.001$ for four different values of $f_{pull}$. The values of $f_{pull}$ are shown in panel (E).
    (E) Fraction of bound linkers in time at the center of the membrane for four different values of $f_{pull}$ with $k_{off}^0/k_{on}=0.001$. $\phi_c$ for $f_{pull}=15 pN$ drops to zero almost instantaneously. 
    (F) Height of the membrane at the center, $z_c$, as a function of the applied force for four different values of $\zeta=k_{off}^0/k_{on}$.
    (G) Fraction of bound linkers at the center of the membrane as a function of the applied force for four different values of $\zeta=k_{off}^0/k_{on}$. The dashed lines correspond to \Cref{eq:phi_eq_1} in the absence of membrane bending and the solid lines with markers are obtained from simulations that couple linker kinetics and membrane deformation.
     In all these simulations, the bending modulus was $\kappa= 90 \ k_BT$, tension was $\lambda = 0.001$ pN/nm, and the linker stiffness was $k_{lin}=0.1$ pN/nm. In panels C and E, the horizontal axis is dimensionless time, where dimensional time is divided by a characteristic time of 4 min to demonstrate the instantaneous nature of the mechanical loading. }
    \label{fig:force_linker}
\end{figure}

While previous studies have focused on linker detachment for small deformations of the membrane or while ignoring membrane bending \cite{Alert2015-zq}, we aimed to study the coupling between membrane bending and linker kinetics in the fully nonlinear regime. 
To do this, we conducted separate simulations of membrane bending due to an applied point load and membrane bending due to a normal pressure acting on the membrane. 
The physical picture is as follows: when an external force (either a point load or a bulk pressure) is applied on the membrane such that the restoring force of the linker springs increases beyond a threshold value, the linkers unbind and the membrane detaches from the cortex (\Cref{fig:force_linker}A).
To understand this behavior, we first focus on the fraction of bound linkers, $\phi$, in the absence of membrane bending. 
From \Cref{eq:phi_eq}, we obtain that the steady state value of $\phi$, in the absence of membrane bending is given by
\begin{equation}
\phi_{eq}=\frac{e^{\frac{-f\delta}{k_BT}}}{\zeta+e^{\frac{-f\delta}{k_BT}}},
    \label{eq:phi_eq_1}
\end{equation}
where $\zeta=k_{off}^0/k_{on}$ and represents the ratio of the two time constants.
For ezrin, the characteristic association time on the membrane was measured to be approximately 6.6 s and the characteristic dissociation time was measured to be 5 s \cite{Fritzsche2014-nh}, suggesting that $\zeta$ is close to 1.
Other experiments note that the dissociation constant of ezrin to PIP2 containing vesicles to be approximately 5 \textmu M \cite{Senju2022-ao, Maniti2013-lb, Blin2008-qe} but the individual rate constants are not available.
To investigate how $\phi_{eq}$ changes with membrane bending, we systematically vary the value of $\zeta$ in our simulations to account for variations in linker proteins and lipid composition along with an applied load on the membrane. 

We first asked whether the relationship between an applied force and membrane bending depends on the linker binding and unbinding time scales.
In other words, when coupled with membrane bending, how does the relationship between $\phi_{eq}$ and force shift?
To answer this question, we simulated the deformation of the membrane in response to a point load of 10 pN for different values of $\zeta$ (\Cref{fig:force_linker}B, C).
First we observed that as the membrane deformed, the regions of high $z$ had zero linkers while the flat regions of the membrane had high values of linkers (\Cref{fig:force_linker}B).
This is as we would expect.
Furthermore, near the neck of the membrane deformation, there was a gradient of linker density. 
We also noted that the final shape of the membrane, for the same fixed force, depended on the value of $\zeta$. 
In the limit of high $\zeta$, the system experiences very weak adhesion and bending dominates for the same value of applied force (see $\zeta=1$, \Cref{fig:force_linker}C).
When $\zeta$ was decreased, such that the unbinding rate was slower than the binding rate, the linkers took longer to unbind and bending membrane was reduced because the adhesion of the linkers resists membrane deformation (\Cref{fig:force_linker}B,C).
Next, we fixed the value of $\zeta$ to 0.001 and varied the pulling force (\Cref{fig:force_linker}D, E). 
As the force increased, a cylindrical tube formed, as expected from previous work \cite{Derenyi2002-ym} and the linker density decreased as the membrane deformation increased (\Cref{fig:force_linker}D).
When the pulling force is large ($f_{pull}=10$ pN and $f_{pull}=15$ pN, \Cref{fig:force_linker}E), we observe an instantaneous detachment of the linkers despite the low value of $\zeta$, indicating that membrane bending dominates at large values of force. 
At lower values of the pulling force, the membrane deformation is small and complete linker detachment is not observed (\Cref{fig:force_linker}E).
The length of the tube, given by $z_c$ is also dependent on $\zeta$ (\Cref{fig:force_linker}F). 
When $\zeta=1$, we obtain a tube length consistent with pure membrane bending \cite{Derenyi2002-ym}.
But as $\zeta$ decreases, the adhesion energy of the linkers increases because of slow unbinding and the resulting tube length is smaller. 
The coupling between linker binding-unbinding and membrane bending also affects $\phi_{eq}$ (\Cref{fig:force_linker}G).
In the absence of membrane bending, the steady state value of $\phi_{eq}$ is given by \Cref{eq:phi_eq_1}. 
This is shown using dashed lines in \Cref{fig:force_linker}F for different values of $\zeta$.
When the linker density is combined with membrane bending due to a pulling force, we see the following behavior: when $\zeta=1$ such that the off rate and on rate are balanced, the coupled model is very close to the simple analytical prediction of steady states. 
When $\zeta$ decreases, such that the linker unbinding is slower than the binding, the force required to reduce the linker density is lower in the coupled model than in the uncoupled model.
Thus, coupling between membrane bending due to a point load and linker kinetics alters the membrane bending landscape.

\begin{figure}[!!h]
    \centering
    \includegraphics[width=0.8\textwidth]{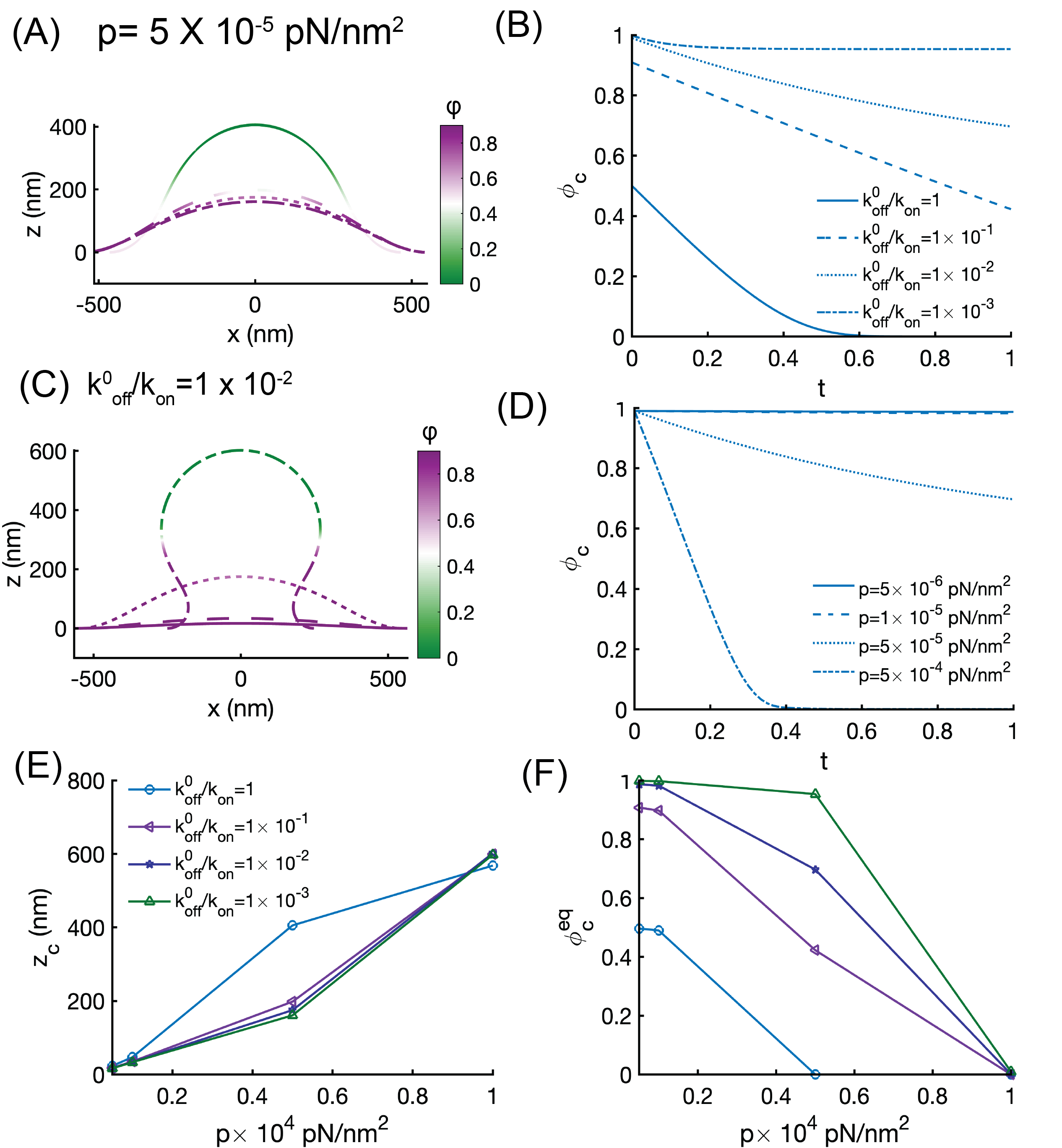}
    \caption{\textbf{Competition between linker binding and pressure determines membrane blebbing.}
    (A) Membrane shapes in response to $p= 5 \times 10^{-5} \ pN/nm^2$ for four different values of $\zeta=k_{off}^0/k_{on}$.
    (B) Fraction of bound linkers at the center of the membrane for four different values of $\zeta=k_{off}/k_{on}$ and $p= 5 \times 10^{-5} pN/nm^2$.
    (C)Membrane shapes in response to $k_{off}/k_{on} = 0.01$ for four different values of pressure.
    (D) Fraction of bound linkers at the center of the membrane for four different values of pressure and for $k_{off}/k_{on}=0.01$.
    (E) Height at the center of the membrane for increasing values of pressure and four different values of $\zeta$.
    (F) Fraction of bound linkers at the center of the membrane as a function of pressure for four different values of $\zeta=k_{off}^0/k_{on}$. The solid lines with markers are obtained from simulations that couple linker kinetics and membrane deformation.
    In all these simulations, the bending modulus was $\kappa= 90 \ k_BT$ tension was $\lambda = 0.001 pN/nm$ and the linker stiffness was $k_{lin}=0.1$ pN/nm. In panels B and D, the horizontal axis is dimensionless time, where dimensional time is divided by a characteristic time of 4 min to demonstrate the instantaneous nature of the mechanical loading. }
    \label{fig:pressure}
\end{figure}

The formation of MPs is often driven by the initiation of a membrane bleb due to fluid pressure \cite{Bhullar2016-rv,Tinevez2009-wn}. 
Therefore, we next investigated how the linker distribution would change due to a pressure difference across the membrane (\Cref{fig:pressure}).
We observed that when a pressure difference was applied across the membrane, the membrane shape tended towards a spherical bud, and the linker density decreased in areas of high deformation (\Cref{fig:pressure}B).
For a fixed value of pressure, the deformation depends on $\zeta$. 
When $\zeta=1$, the linker density decreases over time at the center of the membrane and is accompanied by a larger deformation (\Cref{fig:pressure}A,B).
When the value of $\zeta$ is smaller such that the unbinding of linkers is slower than binding, we see that the membrane deformation is smaller.
This is because the linkers remained bound longer and the adhesion energy associated with these linkers opposes membrane bending. 
Similarly, when we fixed the value of $\zeta$ to 0.01 and varied the applied pressure on the membrane, we found that the linker density decreases faster with increasing pressure (\Cref{fig:pressure}C, D). 
The height of the membrane at the center of the bleb shows a dependence on both the applied pressure and the value of $\zeta$, although the effect of $\zeta$ is smaller than the one observed for a point load (compare \Cref{fig:pressure}E to \Cref{fig:force_linker}E). 
This is likely because a larger area fraction of the membrane is fully detached from the linkers when pressure is applied and bleb formation proceeds through pure membrane bending.
Our simulations predict that the pressure at which the linker density becomes zero also depends on the timescale of linker binding (\Cref{fig:pressure}F). 
When $\zeta$ is small, indicating longer unbinding times, the pressure required to break the linkers increases and when $\zeta=1$ such that binding and unbinding are of comparable times, the linkers are able to completely detach at lower pressure values. 

Thus, our analyses demonstrate that linker detachment is required for membrane bending, and that the timescales of linker binding and unbinding regulate the mechanical response of the membrane to the instantaneous application of either a point load or bulk pressure.
This is consistent with the experimental observations in \cite{Welf2020-hx,Dai1999-fc}.

\subsection*{Interaction between PS kinetics and linker properties determines successful MP formation}
\begin{figure}[!!h]
    \centering
    \includegraphics[width=0.8\textwidth]{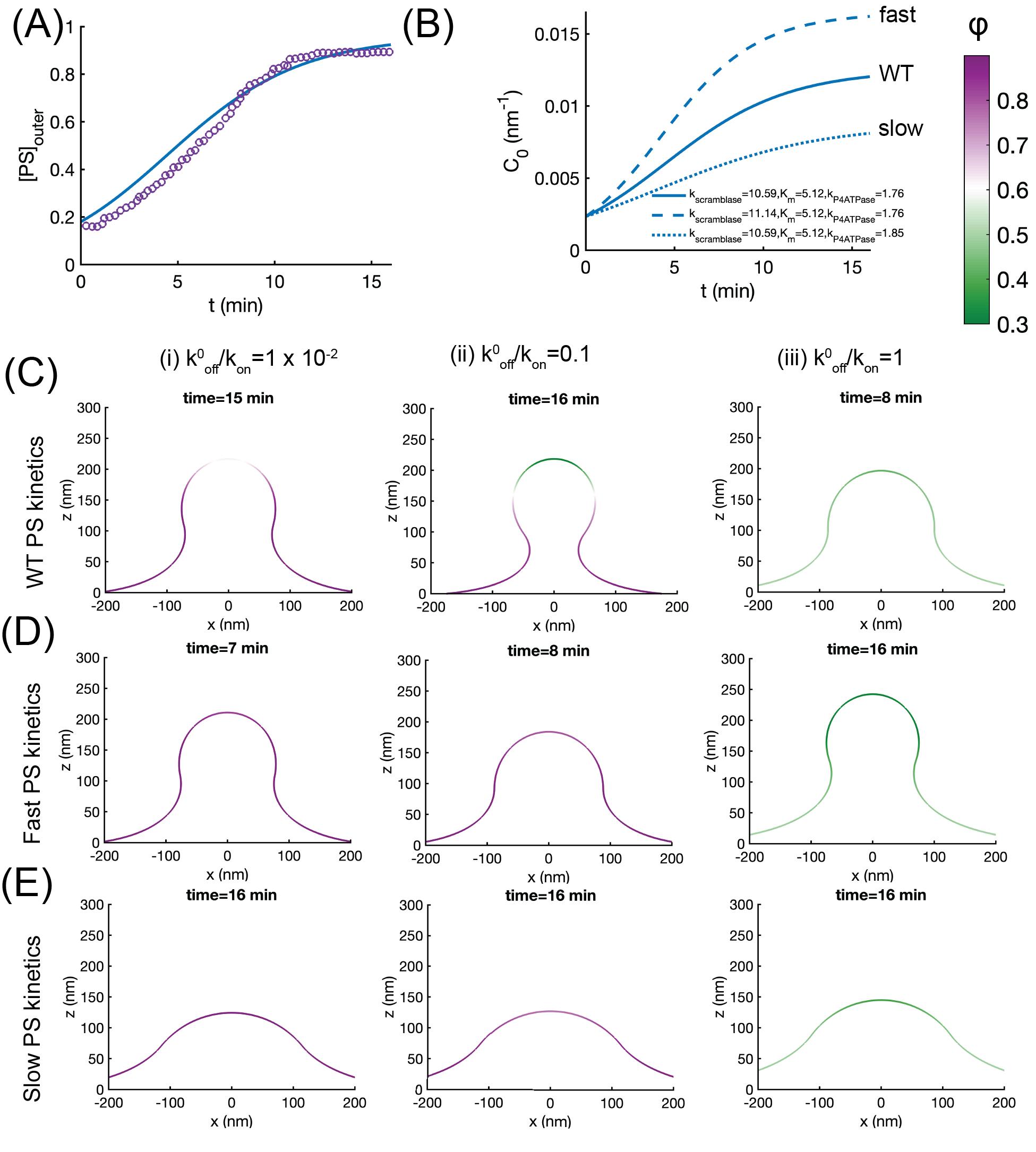}
    \caption{\textbf{MP formation depends on PS kinetics}
    (A) PS kinetics in the model were determined by fitting the rate constants in \Cref{eq:PS_curvature_feedback} to experimental data from \cite{Suzuki2010-pp} to obtain a time course for PS kinetics.
    (B) Spontaneous curvature functions used in our model. WT kinetics refers to the time course that matches PS kinetics in panel (A), slow and fast kinetics are obtained by modifying the kinetic parameters in our model as shown in the legend.
    (C) Membrane shapes for WT PS kinetics for (i) $\zeta=0.01$, (ii) $\zeta=0.1$, (iii) $\zeta=1$. 
(D) Membrane shapes for fast PS kinetics  for (i) $\zeta=0.01$, (ii) $\zeta=0.1$, (iii) $\zeta=1$. 
   (E) Membrane shapes for slow PS kinetics for (i) $\zeta=0.01$, (ii) $\zeta=0.1$, (iii) $\zeta=1$.  
   The times at which a bud formed are shown for each panel and depend on the PS kinetics. 
     In all these simulations, the bending modulus was $\kappa= 90 k_BT$, tension was $\lambda = 0.001$ pN/nm, and spontaneous curvature area was 50265 nm$^2$. Linker stiffness was maintained at $k_{lin}=0.05$ pN/nm.
     The color bar shows the fraction of bound linkers along the membrane stiffness.}
    \label{fig:PS_kinetics}
\end{figure}

Under conditions of inflammation, PS flipping to the outer leaflet is the key step for MP formation.
The kinetics of PS flipping depend on the kinetics of the scramblase and the P4ATPase (\Cref{fig:MicroparticleSketch}A, \Cref{eq:PS_curvature_feedback}).
Experiments use Annexin V as a readout for the presence of PS on the outer surface \cite{Suzuki2010-pp}. 
In order to fit the rate parameters for our model, we used experimental data from \cite{Suzuki2010-pp} because this study focuses primarily on the role of TMEM16F.
We fit our model with the data from Figure 2B of \cite{Suzuki2010-pp} to obtain the rate parameters for the wild type (WT) condition in our model (\Cref{fig:PS_kinetics}A).
Since these values are in arbitrary units, we map the PS kinetics to a spontaneous curvature value as a function of time as shown in \Cref{fig:PS_kinetics}B.
Based on experimental observations of annexin V staining \cite{Suzuki2010-pp} we used a time scale of approximately 15-16 min.
We also vary the kinetics of PS-dependent spontaneous curvature to be faster than the WT or slower than the WT to investigate the effect of scramblase dominance and P4ATPase dominance, respectively.
For interpretation, we define successful MP formation when the membrane shape develops an overhang (a U-shape such that the angle of the tangent at the neck with the horizontal is greater than $\pi/2$; see supplementary material).
We simulated three different PS kinetics, the first tuned to mimic WT PS kinetics, and also a slow and a fast case to mimic misregulation of PS exchange between the leaflets, without (\Cref{fig:PS_kinetics}) and with curvature-coupled feedback (\Cref{fig:curvature_feedback}).
For each of these cases we used three different values of $\zeta$ to identify how coupling PS kinetics (and the resulting changes in preferred curvature) with linker kinetics impacts successful MP formation and shape.

Using experimentally-constrained WT PS kinetics, without any curvature-coupled feedback (\textit{i.e.} $\alpha$ and $\beta$ are zero), we note that for all values of $\zeta$, the shape of the membrane shows a well-formed bud, consistent with the effect of increased PS-induced spontaneous curvature (\Cref{fig:PS_kinetics}C).
Interestingly, we observed that the shape of the membrane, particularly the diameter of the neck, depends on the particular value of $\zeta$. 
For $\zeta=0.01$ and $\zeta=1$, the membrane neck is wide even though the density of bound linkers is different for these two cases (\Cref{fig:PS_kinetics}Di, iii). 
For $\zeta=0.1$, a well-formed neck with a catenoid-like shape is observed (\Cref{fig:PS_kinetics}Dii). 
When we simulate the membrane shapes with fast PS kinetics, we observe that while a successful MP can form faster, the neck remains wider for all values of $\zeta$ (\Cref{fig:PS_kinetics}D).
For slow PS kinetics across the three different values of $\zeta$ tested, we found that the shape of the membrane remains a shallow protrusion at 16 min (\Cref{fig:PS_kinetics}E).
Based on these simulations, \Cref{fig:PS_kinetics}C-E) map a phase diagram for the interaction between linker binding properties and PS kinetics.
The effect of curvature-coupled feedback on the shapes of the membrane was weak, possibly because of the simplified model we used (\Cref{fig:curvature_feedback}).
Additionally, membrane tension and bending modulus also alter the MP formation (\Cref{fig:membrane_properties}), consistent with other tension-dependent membrane processes \cite{Hassinger2017-su, Alimohamadi2023-ep, Rangamani2022-bv}.
This is consistent with experimental observations showing that imbalance of scramblase-flippase activity results in reduced MP formation, which can dampen the procoagulation cascade \cite{Morel2011-wf,Suzuki2010-pp}.
These results indicate that there exists a balance between the rate of spontaneous curvature induction through PS flipping and linker binding and unbinding.
Loss of this balance due to disease states \cite{Morel2011-wf,Perez-Cornejo2012-gy} can alter mechanics, leading to a failure in MP formation. 
These simulations predict that the equilibrium shape of the membrane, and therefore successful MP formation, not only depends on the spontaneous curvature induced by the PS, as expected, but also on the interplay between linker lifetimes and the kinetics of PS flipping.

\begin{figure}[!!h]
    \centering
    \includegraphics[width=0.8\textwidth]{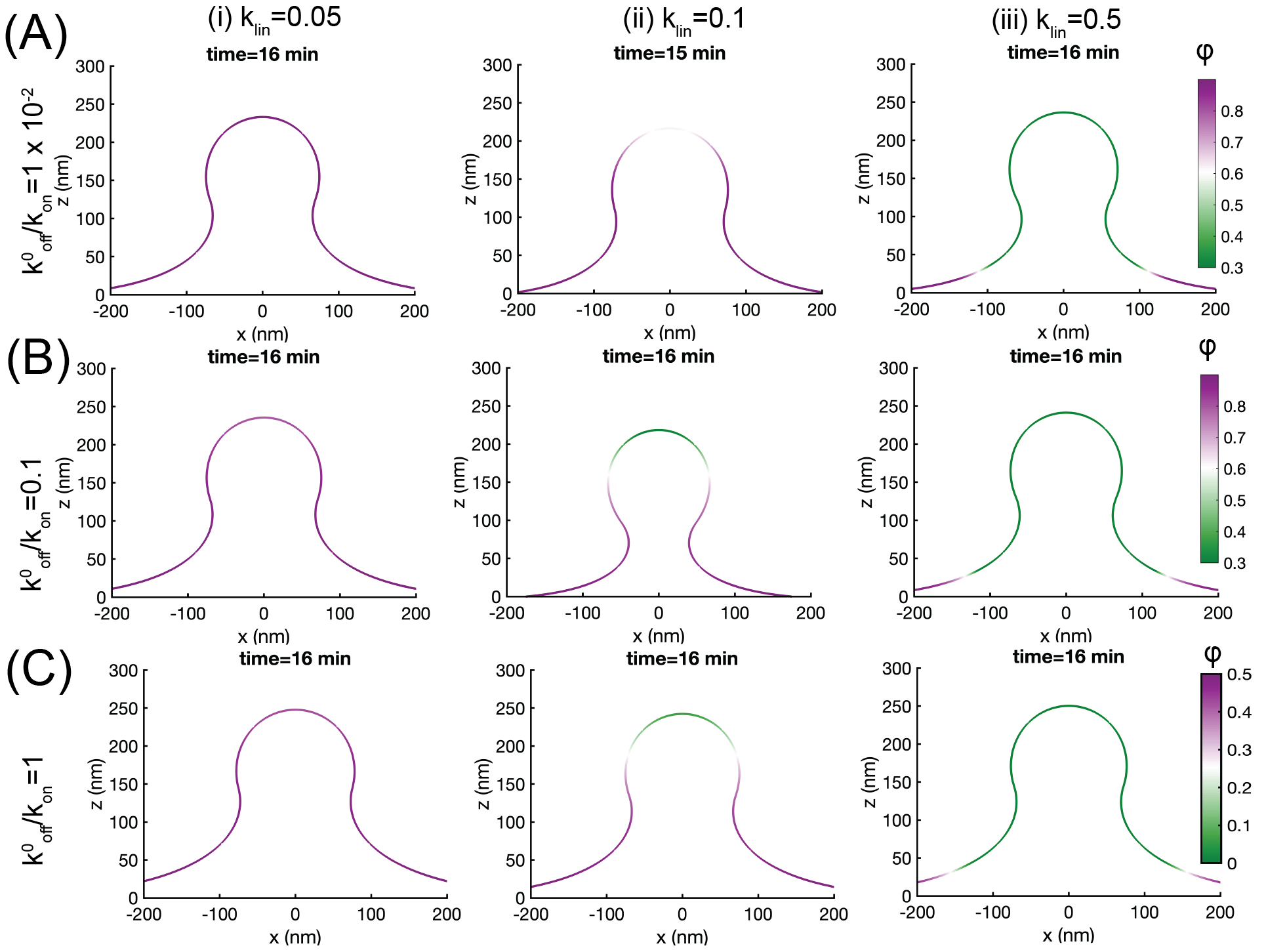}
    \caption{\textbf{Linker stiffness determines the effect of PS-induced spontaneous curvature for MP formation. }
    (A) Membrane shape for $\zeta=0.01$, WT PS kinetics, and three different values of linker stiffness: (i) $k_{lin}=0.05$ pN/nm, (ii) $k_{lin}=0.1$ pN/nm, (iii) $k_{lin}=0.5$ pN/nm.
    (B) Membrane shape for $\zeta=0.1$, WT PS kinetics, and three different values of linker stiffness: (i) $k_{lin}=0.05$ pN/nm, (ii) $k_{lin}=0.1$ pN/nm, (iii) $k_{lin}=0.5$ pN/nm.
      (C) Membrane shape for $\zeta=1$, WT PS kinetics, and three different values of linker stiffness: (i) $k_{lin}=0.05$ pN/nm, (ii) $k_{lin}=0.1$ pN/nm, (iii) $k_{lin}=0.5$ pN/nm.
    The times at which a bud formed are shown for each panel and depend on the PS kinetics. 
     In all these simulations, the bending modulus was $\kappa= 90 k_BT$, tension was $\lambda = 0.001$ pN/nm, and spontaneous curvature area was 50265 nm$^2$.
     The color bar shows the fraction of bound linkers along the membrane.}
    \label{fig:linker_stiffness}
\end{figure}

In addition to the binding and unbinding rates, another important linker property is the spring stiffness, $k_{lin}$.
While the exact value of this stiffness is not known, it is possible that the linker crosstalk with the cytoskeleton \cite{Korkmazhan2022-xs}, and with the scramblases \cite{Perez-Cornejo2012-gy} could lead to altered stiffness. 
We next asked whether the linker stiffness could alter the energy landscape of successful MP formation. 
For all the cases tested, we used WT PS kinetics and varied $\zeta$ and $k_{lin}$ across a range of parameters.
In all cases, increase $k_{lin}$ for a given value of $\zeta$ leads to an increase in the fraction of unbound linkers.
For low values of $\zeta=0.01$, we found that increasing linker stiffness leads to a well-formed neck (compare \Cref{fig:linker_stiffness}Ai, ii, and iii).
However, for $\zeta=0.1$, increasing linker stiffness first led to a better formed neck (compare \Cref{fig:linker_stiffness}Bi and ii) but further increase led to neck widening (compare \Cref{fig:linker_stiffness}Bii and iii).
For $\zeta=1$, no observable changes to the membrane shape or neck are seen for the different values of $k_{lin}$.
Thus, we predict that there exists a complex feedback between linker properties and membrane properties.
We observed that the shape of the membrane also depends on the value of the linker stiffness in both cases, highlighting the feedback between membrane mechanics and linker stiffness.

\subsection*{Combining external load with PS kinetics leads to faster MP formation}
\begin{figure}[!!h]
    \centering
    \includegraphics[width=\textwidth]{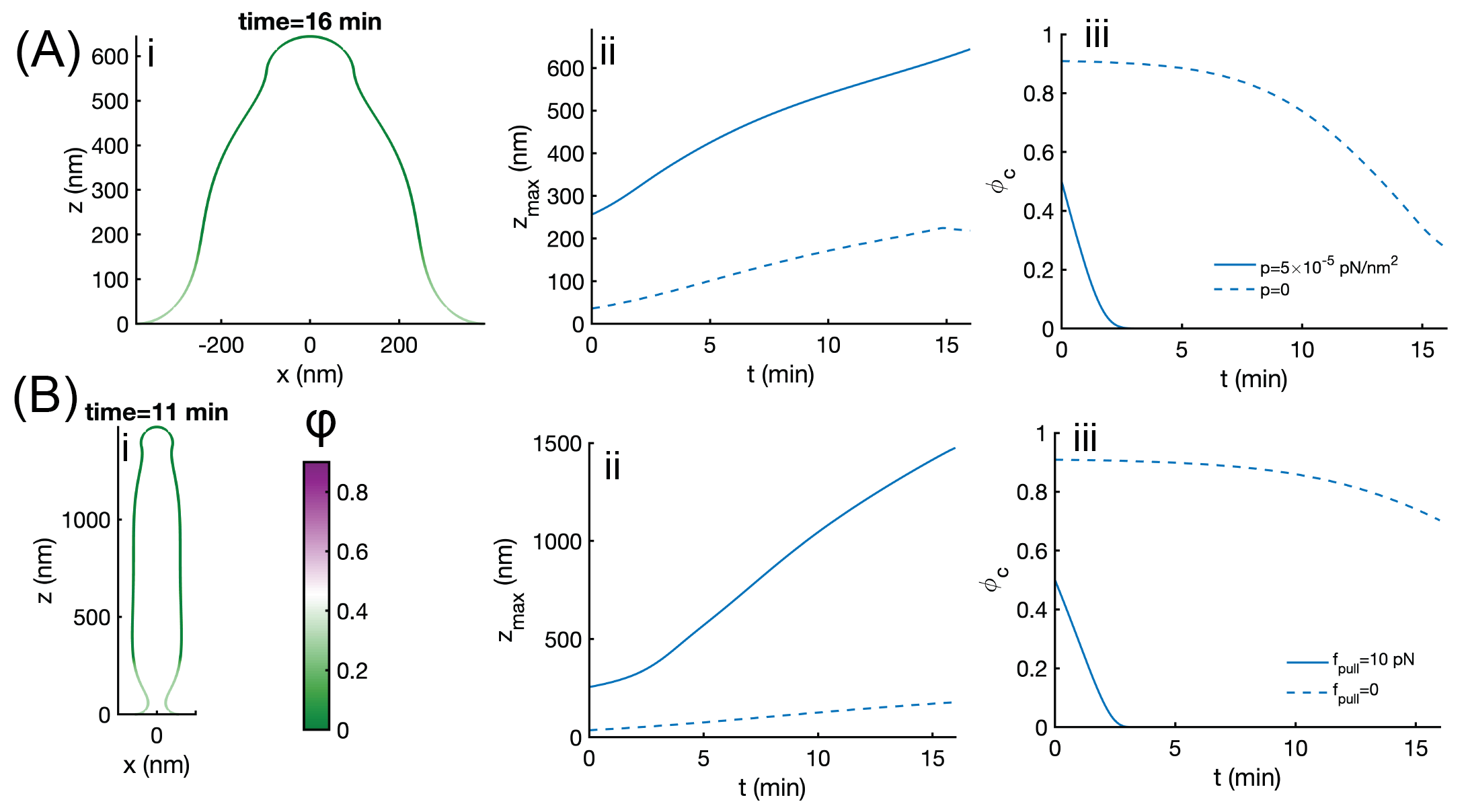}
    \caption{\textbf{Combination of membrane parameters and external forces aid linker detachment and promote MP formation. $\zeta=1$}
    (A) Effect of pressure $5\times10^{-5}$ in the presence of PS-induced spontaneous curvature (WT kinetics) on (i) equilibrium shapes, (ii) temporal evolution of membrane deformation ($z_{max}$), and (iii) temporal evolution of linker density at the center of the membrane. 
    (B) Effect of localized force 10 pN in the presence of PS-induced spontaneous curvature (WT kinetics) on (i) equilibrium shapes, (ii) temporal evolution of membrane deformation ($z_{max}$), and (iii) temporal evolution of linker density at the center of the membrane.
    In (A) and (B), the bending modulus was $\kappa= 90 k_BT$, tension was $\lambda = 0.001$ pN/nm, $k_{lin}=0.1$ pN/nm, and spontaneous curvature area 50265 nm$^2$.
    }
    \label{fig:combination}
\end{figure}

Thus far, our simulations have shown that membrane deformation and unbinding from linkers is quite effective when an instantaneous pressure or force is applied to the membrane (\Cref{fig:force_linker,fig:pressure}) whereas when only PS-induced spontaneous curvature is considered, the membrane deformation depends on the kinetics of PS flipping and linker properties.
In cells, when MPs form, a number of these factors are thought to act together -- PS flipping occurs over a long time scale \cite{Suzuki2010-pp} but blebs and MPs can form as quickly as in 30 s \cite{Charras2006-gr} potentially because the time scale for pressure or force-induced deformation is shorter.
Therefore, we investigated what happens when external loading such as pressure or localized forces are applied to a membrane that has linker adhesion and PS-induced spontaneous curvature. 
When we used a pressure (\Cref{fig:combination}A) or localized force (\Cref{fig:combination}B) with PS WT kinetics, we observed the following: the membrane deformation is larger in either case than just external loading or spontaneous curvature. This is to be expected because of the stresses acting on the membrane are larger when both sources of membrane bending are present (\Cref{fig:combination}Ai, Bi). 
The dynamics of the deformation is significantly faster for pressure or force is present when compared to just PS-induced spontaneous curvature (\Cref{fig:combination}Aii, Bii).
We also observed that the linker deformation is faster and the linkers are completely detached from the membrane in these conditions (\Cref{fig:combination}Aiii, Biii).
Thus, our simulations reveal that combination of external loading with PS kinetics can lead to increased linker detachment, rapid and larger MP formation.

\section*{Discussion}
\begin{figure}[!!h]
    \centering
    \includegraphics[width=0.9\textwidth]{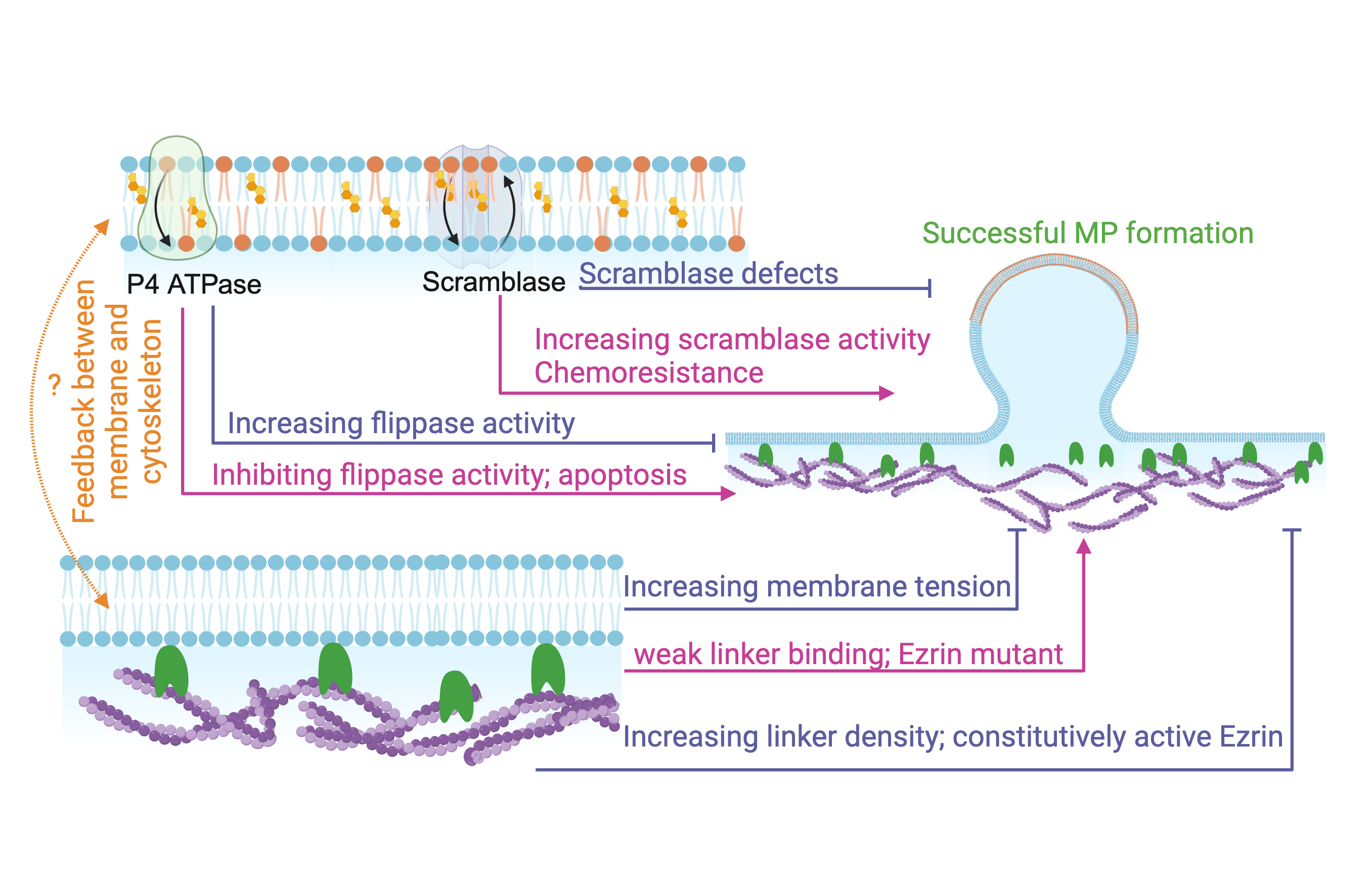}
    \caption{\textbf{Interplay between linker properties and lipid properties determines appropriate MP formation.}
  Successful MP formation is a balance of lipid properties and linker strength. Loss of this balance through disease states and drug-induced states can lead to the overproduction or underproduction of MPs. The schematic summarizes a few scenarios that could play a role and outlines the need to explore feedback between the different biophysical players to further shed light into this critical membrane budding phenomenon. }
    \label{fig:discussion}
\end{figure}

Curvature generation in cell membranes is an important biophysical phenomenon that can be explained well by mechanical arguments \cite{Helfrich1973-lr,Hassinger2017-su,Mahapatra2020-dp,Malingen2022-hy}.
In this study, we focused on the formation of MPs as a example to study how outward buds can be generated from the plasma membrane. 
To pull away from the cytosol, membrane deformation due to spontaneous curvature or pressure requires detachment from the cortex \cite{Welf2020-hx,Dai1999-fc}. 
Our model leads to the following predictions: the strength of attachment of the membrane to the cortex, determined by a combination of linker stiffness and the binding-unbinding rates of linkers, determines the success of MPs or blebs from the cell membrane.
The interplay between linker binding and PS-induced spontaneous curvature suggests a Goldilocks principle for MP formation: very strong or very weak linkers are not conducive to membrane-cortex detachment; very fast or very slow kinetics of PS-induced spontaneous curvature are not conducive to MP formation.
A balance between membrane-cortex linkers and membrane-curvature is essential for forming MPs. 
Many factors can disrupt this balance in the case of disease states, chemoresistance, and drug treaments and our models suggest ways in which opposing factors can be combined to either promote or mitigate MP formation depending on the physiological or pathological state (\Cref{fig:discussion}).

A major conclusion from our model is that in the presence of linkers factors such as PS-induced spontaneous curvature generation and modulation of membrane properties including bending modulus and tension are necessary but not sufficient to ensure adequate curvature generation of the membrane (across the physiologically relevant values we tested).
These findings are consistent with experiments under different conditions that show that ezrin is not present in an expanding bleb \cite{Charras2006-gr} and that linker detachment is necessary for membrane protrusions \cite{Welf2020-hx}.
Based on our predictions, modulating the linker properties should then modify the propensity to form blebs.
Charras and colleagues \cite{Charras2006-gr} showed that mutant forms of ezrin modulate the attachment of the plasma membrane to the cortex. 
Constitutively active ezrin, which can be expressed as strong binding (low $\zeta$) leads to loss of blebbing, while expressing only the FERM domain weakens the membrane-cortex attachment (See Figure 8 of \cite{Charras2006-gr}).

The other major finding from our model is that the rate of spontaneous curvature generation induced by PS flipping kinetics can play an important role in determining the crosstalk between linker detachment and membrane curvature generation. 
The extent of MP dependence on PS flipping is quite relevant to a disease state known as Scott syndrome \cite{Morel2011-wf}.
Scott syndrome is characterized by decreased MP formation (and a resulting decrease in procoagulant activity) because of an inherited defect in lipid scramblase activity \cite{Morel2011-wf,Piccin2007-js,Suzuki2010-pp}.
Studies of the lipid scramblase of the TMEM family have shown that in addition to lipid scrambling, these calcium-activated ion channels interact stoichiometrically with the ezrin-radixin-moesin network \cite{Perez-Cornejo2012-gy} and can be negatively regulated by the cytoskeleton \cite{Lin2018-jm}. 
While our model simplifies the PS-induced spontaneous curvature, we are able to show that slow PS kinetics (scramblase defect) and fast PS kinetics (constitutively active scramblase) play a role in MP formation. 
Our findings also have implications for understanding the mechanisms of chemoresistance. 
In cancer cells, MPs shed from the cell membrane are used to efflux the drug out of the cancer cells. 
It has been suggested that inhibition of externalization of PS using bisindolylmaleimide-I and d-pantethine, could be used in conjunction with chemotherapy \cite{Hayatudin2021-ph}.

Our model only focused on membrane deformation and linker attachment but not changes to cytosolic fluidity, calcium signaling, in-plane rearrangement of lipids and proteins in the plasma membrane, and the associated cytoskeleton remodeling for computational simplicity.
Including these processes is an important aspect of tying the mechanistic aspects of MP formation to the biophysical models and will be a focus of future studies. 
Despite these simplifications, we find that our model is able to shed light on some of the primary aspects of outward budding of the cell membrane in processes such as blebbing and microparticle formation.

\section*{Acknowledgements}
We thank the members of the Rangamani lab for their feedback and Profs. Itay Budin (UCSD) and Jeanne Stachowiak (UT Austin) for extensive discussions.
This project was supported by Office of Naval Research N00014-20-1-2469 to P.R.

\section*{Author contributions}
SAM and PR conceptualized the project scope and designed the simulations, AM derived the continuum description and conducted simulations. All authors co-wrote the manuscript. Funding was obtained by PR.

\clearpage
\begin{centering}
    \textbf{\Large Supplementary Material for\\
    Interplay between cortical adhesion and membrane bending regulates microparticle formation}\\[3mm]
    \textbf{A. Mahapatra$^{1, 2}$, S. A. Malingen$^{1,4}$, and P. Rangamani$^{1,**}$}\\[1mm]
    $^{1}$ Department of Mechanical and Aerospace Engineering,\\
    University of California San Diego, La Jolla CA 92093.\\
    $^{2}$ Present address: Senior Research and Development Engineer, Gamma Technologies, LLC, Westmont, IL  \\
    $^{4}$ Present address: Department of Bioengineering, University of Washington, Seattle \\
    $^{**}$To whom correspondence must be addressed: prangamani@ucsd.edu\\
\end{centering}

\setcounter{figure}{0}
\renewcommand{\figurename}{Fig.}
\renewcommand{\thefigure}{S\arabic{figure}}
\setcounter{equation}{0}
\renewcommand{\theequation}{S\arabic{equation}}


\section*{Model Development}
\label{sec:si_sec1}
\subsection*{Surface representation}
\label{sec:si_sec1_1}
\begin{figure}[!!h]
    \centering
    \includegraphics[width=\textwidth]{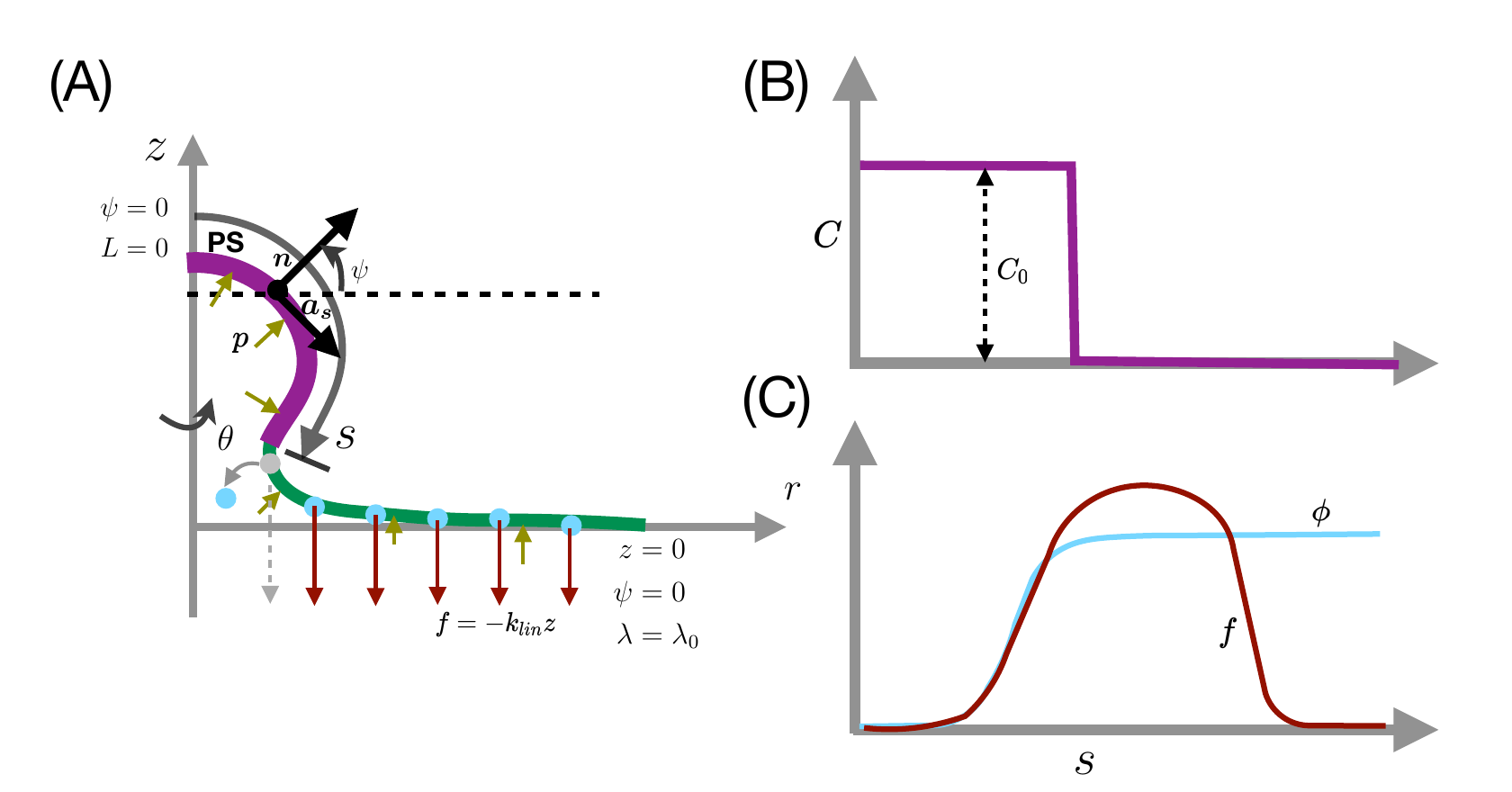}
    \caption{\textbf{Schematic of membrane configuration with inputs and boundary condition.}
  (A) Simulation domain of the membrane, distribution of applied force and spontaneous curvature and boundary conditions. (B) Distribution of spontaneous curvature along the arclenth of the membrane. (C) Distribution of linker density and resultant applied force on the membrane along the arclength of the membrane.}
    \label{fig:schematic_supp}
\end{figure}

In polar coordinates, we can parameterize the membrane by the arclength $s$ and the rotation angle $\theta$ (\Cref{fig:schematic_supp}a) as
\begin{equation}
    \boldsymbol{r}(r,z,\theta)=\boldsymbol{r}(s,\theta).
\end{equation}
The surface tangents are given by $\boldsymbol{e}_s=\boldsymbol{r}_{,s}$ and $\boldsymbol{e}_\theta=\boldsymbol{r}_{,\theta}$.
The surface metric $a_{\alpha \beta}=\boldsymbol{e}_\alpha \cdot \boldsymbol{e}_{\beta}$ and the curvature tensor $b_{\alpha \beta}=\boldsymbol{e}_{\alpha,\beta}\cdot \boldsymbol{n}$ are the two fundamental tensors we use in the derivation that follows. We refer the reader to \cite{Steigmann1999-es,Kreyszig1969-hv} for details of these derivations. The scalar invariants mean curvature $H$ and Gaussian curvature $K$ denote average and product of the two principal curvatures and are given by

\begin{equation}
H=\frac{1}{2} a^{\alpha \beta} b_{\alpha \beta}, \quad K=\frac{1}{2} \varepsilon^{\alpha \beta} \varepsilon^{\mu \eta} b_{\alpha \mu} b_{\beta \eta},
\end{equation}
where, $a^{\alpha \beta}=\left(a_{\alpha \beta}\right)^{-1}$, and $\varepsilon^{\alpha \beta}$ is the permutation tensor: $\varepsilon^{12}=-\varepsilon^{21}=1 / \sqrt{a}, \varepsilon^{11}=$ $\varepsilon^{22}=0$, with $a=\det|{a_{\alpha \beta}}|$.
Here $\alpha,\beta$ are the surface coordinates, and moving forward we use Greek letters to indicate surface coordinates.

\subsection*{Force balance}
The force balance equation is dictated by 
\begin{equation}
\label{eq:stress_bal}
 \boldsymbol{T}^{\alpha}_{;\alpha}+p\boldsymbol{n}+\boldsymbol{f}=0\,,
\end{equation}%
where $p$ is normal pressure on the membrane, $()^\alpha_{;\alpha}$ is the surface divergence, and $\boldsymbol{T}$ is thetraction on the membrane and given by,
\begin{equation}
\boldsymbol{T}^{\alpha }=N^{\beta \alpha }\boldsymbol{a}_{\beta }+S^{\alpha }\boldsymbol{n}.
\label{eqn:stress_balance_2}
\end{equation}%
$\boldsymbol{N}$ is the in-plane component of the stress and is given by
\begin{equation}
 N^{\beta \alpha }=\zeta^{\beta \alpha }+b_{\mu }^{\beta }M^{\mu
\alpha} \qquad \mathrm{and} \qquad S^{\alpha }=-M_{;\beta }^{\alpha \beta },
\label{eqn:stresses}
\end{equation}
where $\sigma^{\beta \alpha}$ and $M^{\beta \alpha}$ are obtained from the following constitutive relations \cite{Steigmann1999-es}
\begin{equation}
   \sigma^{\beta \alpha }=\rho \left(\frac{\partial F}{\partial a_{\alpha \beta}}+\frac{\partial F}{\partial a_{\beta \alpha}}\right) \quad \text{and} \quad \boldsymbol{M}^{\beta \alpha }=\frac{\rho}{2} \left(\frac{\partial F}{\partial b_{\alpha \beta}}+\frac{\partial F}{\partial b_{\beta \alpha}}\right),
\end{equation}
with $F=W/\rho$ as the energy mass density of the membrane.
Combining these we get the balance equations in \Cref{eq:stress_bal} in normal direction
\begin{align}
\label{eqn:norm_gen}
p+\boldsymbol{f} \cdot \mathbf{n}=-(S^{\alpha}_{;\alpha}+N^{\beta \alpha}b_{\beta \alpha})&=\Delta\left(\frac{1}{2} W_H\right)+\left(W_K\right)_{; \alpha \beta} \tilde{b}^{\alpha \beta}+W_H\left(2 H^2-K\right)\\
&+2 H\left(K W_K-W\right)-2 \lambda H,
\end{align}
with $\tilde{b}^{\alpha \beta}=2a a^{\alpha \beta}-b^{\alpha \beta} $ and in the tangential direction as
\begin{equation}
\label{eqn:tangen_gen}
-\boldsymbol{f}\cdot \boldsymbol{\tau}=N_{; \alpha}^{\beta \alpha}-S^\alpha b_\alpha^\beta=a^{\beta \alpha}\left(\frac{\partial W}{\partial \theta^\alpha_{|exp}}+ \lambda_{,\alpha}\right),
\end{equation}
where ${\partial \theta^\alpha_{|exp}}$ denotes the explicit coordinate dependence of the free energy.

%

\subsubsection*{Governing equations in axisymmetry and numerical implementation }
We solved the governing equation \Cref{eqn:norm_gen}-\ref{eqn:tangen_gen} in the limit of axisymmetry where $\frac{\partial}{\partial \theta}\approx 0$. The system of equations can be represented in terms of arclength $s$ as shown in \Cref{fig:schematic_supp}a.
The surface metric and curvature tensor in axisymmetry becomes
\begin{equation}
a_{\alpha \beta}=\left[\begin{array}{ll}
1 & 0 \\
0 & r^{2}
\end{array}\right],
\end{equation}
and
\begin{equation}
b_{\alpha \beta}=\left[\begin{array}{cc}
\psi_s & 0 \\
0 & r \sin \psi
\end{array}\right].
\end{equation}
The mean curvatures and Gaussian curvature simplify as
\begin{equation}
H=\frac{1}{2}\left( \psi_s +\frac{\sin \psi}{r}\right), \quad K=\psi_s \frac{\sin \psi}{r}.
\end{equation}
Note that we introduced an angle $\psi$, which is the angle made by the tangent of the membrane with the cortex plane (see \Cref{fig:schematic_supp}a).
The tangential and normal components of the adhesive force becomes function of $\psi$ as:
\begin{equation}
\boldsymbol{f}=f_s \boldsymbol{e}_s+f_n \boldsymbol{n}
\quad \text{where, } 
    f_s=f\sin \psi=-k_{\mathrm{lin}} z \sin \psi
\quad \text{and }
    f_n=f\cos \psi=-k_{\mathrm{lin}} z \cos \psi. 
\end{equation}
Note that, the total adhesive cortex force $\boldsymbol{f}=f_s \boldsymbol{e}_s+f_n \boldsymbol{n}$. Using the free energy given in \Cref{eqn:free_energy}, the tangential force balance along the arclength becomes \cite{Steigmann1999-es,Hassinger2017-su,Mahapatra2020-dp}
\begin{equation}\label{eqn:tangentialbalance}
\frac{\partial \lambda}{\partial s} +f_s  =\frac{\partial C}{\partial s}\left[2 \kappa (H-C)\right].
\end{equation}
Here, $\lambda$ is the Lagrange multiples for area extensibility, and often interpreted as membrane tension \cite{Rangamani2014-tq,Rangamani2022-bv}.
On the other hand, the normal force balance illustrates shape of the membrane, and is given by \cite{Steigmann1999-es,Hassinger2017-su,Mahapatra2020-dp}
\begin{equation}
\label{eqn:normalbalance}
\begin{aligned}
 \frac{1}{r} & \frac{\partial}{\partial s}\left(r \frac{\partial( \kappa (H-C))}{\partial s}\right)+2 \kappa (H-C)\left(2 H^{2}-K\right) \\
&-2 H\left[\kappa(H-C)^{2}\right] =p+f_n+2 \lambda H.
\end{aligned}
\end{equation}
In \Cref{eqn:normalbalance}, $p$ is the normal pressure applied to the membrane and $f_n$ is the normal component of the cortex force on the membrane. 

We solved the governing equation (\cref{eqn:tangentialbalance}-\ref{eqn:normalbalance}) 
numerically in an axisymmetric membrane with temporally evolving spontaneous curvature at according to \cref{eq:PS_curvature_feedback}  
and the spatiotemporal density of linkers across the simulation domain. The solution domain and boundary condition is presented in \Cref{fig:schematic_supp}. 
We split the higher-order shape equation into two lower-order equations.
We define $L$ such that
\begin{equation}
    \frac{L}{r}=\frac{\partial (\kappa (H-C))}{\partial s},
\end{equation}
and use it to express the normal force: 
\begin{equation}
\label{eqn:normal_ODE}
\begin{aligned}
 &\frac{1}{r} \frac{\partial L}{\partial s} +2 \kappa (H-C)\left(2 H^{2}-K\right) -2 H\left[\kappa(H-C)^{2}\right] =p+f_n+2 \lambda H.
\end{aligned}
\end{equation}
Notably, $L$ can be interpreted as the normal bending stress on the membrane, and we used the boundary condition of $L=0$ at the center of the membrane, indicating zero point force at the pole.
To solve the tangential and normal force balance equation, we used the MATLAB-based boundary value solver (bvp4c).
The kinetic equation of spontaneous curvatures and linkers density is solved with explicit time marching. 
We used a nonuniform grid ranging from 5,000 to 100,000 points across the radial domain, with the finer grid towards the pole. We used a large solution domain size of $10^4$ nm$^2$ compared to MP area to avoid boundary effects.

Ultimately, we solve a system of six ODEs for the membrane shape. 
Three of these equations arise from the assumption of axisymmetry: 
\begin{equation}
    r' = \cos \psi \quad
    z' = \sin \psi \quad
    r\psi' = 2rH-\sin\psi 
\end{equation}
Finally, we also solve Equations \ref{eqn:tangentialbalance} and \ref{eqn:normal_ODE}.
We prescribe the following boundary conditions at the pole $s = 0$:
\begin{equation}
r(0^{+}) = 0, \ L(0^{+}) = 0, \ \psi (0^{+}) = 0.
\end{equation}
At the edge ($s = S$) we prescribe the boundary conditions: 
\begin{equation}
z(S) = 0, \ \psi(S) = 0, \ \lambda (S) = \lambda_{0}.
\end{equation}
\subsection*{Pulling force on the membrane}
The pulling force on the membrane depends on the density of the linkers protein on the membrane and the vertical displacement from the cortex plane. Note that each linker works as a linear spring and applies a force in the z-direction of $f_z=\mathbf{f}\cdot \mathbf{z}=-k_{\text {lin }} z$. With $\phi$ being the area fraction of the linker density, the force per unit area $\boldsymbol{f}=\phi f_z \hat{\mathbf{z}}$. The area fraction is obtained by solving the binding and unbinding equation for linker density in time as given by \Cref{eq:phi_eq}.

\subsection*{Spontaneous curvature}
The spontaneous curvature on the other-hand depends on the density of PS on the outer leaflet of the membrane which is governed by \Cref{eq:PS_curvature_feedback}. 
The area of lipid for PS, $a_{PS}$, is around 58 $nm^2$ and and induces a radius of curvature on the membrane about $14.4~nm$. Therefore the spontaneous curvature of each PS $C_0\approx 0.07~nm^{-1}$ \cite{Fuller2003-ws}. 
We use the PS kinetics from experiments to propose a spontaneous curvature function as 
$C([PS])=\gamma [PS]$, where $\gamma=C_0 a_{PS}$.
Multiplying the PS kinetics by this constant factor $\gamma$ gives us the rate equation for spontaneous curvature. 
\section*{Supplementary Tables}
The notation \Cref{table:Notation} and values of parameters \Cref{table:parameters} are summarized below.

\begin{table}
\caption{Notation used in the model}
\label{table:Notation}
\centering
\begin{tabular}{lll}
\hline Notation & Description & Unit \\
\hline
   $W$ & Free energy density per unit area of the membrane  &  $\mathrm{pN \cdot nm ^{-1}}$ \\
  $\kappa$ & Bending modulus of the membrane  & $\mathrm{pN \cdot nm}$ \\
  $ s$ & Arclength along the membrane & $\mathrm{nm}$ \\
  $ \psi$ & Angle made by surface tangent with the horizontal direction  & 1 \\
$H$ & Mean curvature & $\mathrm{nm}^{-1}$ \\
$C$ & Spontaneous mean curvature & $\mathrm{nm}^{-1}$ \\
$\boldsymbol{f}$ & Adhesion force per unit area between the membrane and the cortex & $\mathrm{pN \cdot nm ^{-2}}$ \\
$k_{lin}$ & linear spring constant of the linker & $\mathrm{pN \cdot nm ^{-1}}$ \\
  $p$ & Normal pressure acting on the membrane & $\mathrm{pN \cdot nm ^{-2}}$ \\
$\lambda$ & Membrane tension & $\mathrm{pN \cdot nm ^{-1}}$ \\
$k_{on}$ & Association rate constant of linkers  & $\mathrm{ sec ^{-1}}$ \\
$k_{off}$ & Dissociation rate constant of linkers  & $\mathrm{ sec ^{-1}}$ \\
\hline
\end{tabular}
\end{table}

\begin{table}
\caption{List of parameters}
\label{table:parameters}
\centering
\begin{tabular}{lll}
\hline Notation & Description & Range \\
\hline
 $\kappa$ & Bending modulus of the membrane  & $40-600~\mathrm{pN\cdot nm}~(10-150 ~ \mathrm{k_BT})$  \cite{Monzel2016-ki} \\ 
$\lambda_0$ & Membrane tension at the boundary & $0.001 \textendash 0.1 ~\mathrm{pN \cdot nm ^{-1}}$  \cite{Lipowsky2013-py,Rangamani2022-bv} \\
$a_{\mathrm{coat}}$ & Coat area of the PS domain based on MP sizes & $2.5 \times 10^{4} - 5 \times 10^{4} ~\mathrm{nm}^{2}$ \cite{Piccin2007-js} \\ 
$k_{\mathrm{lin}}$ & Linear spring constant of cortex & $0.01  - 0.1 ~\mathrm{pN \cdot nm}^{2}$ \cite{Alert2015-zq} \\ %
$k_{on}$ & Association rate constant of the linkers & $0.01  - 0.1 ~\mathrm{ sec ^{-1}}$ \cite{Alert2015-zq,Tsai2018-op,Fritzsche2014-nh}\\
$k_{off}$ & Dissociation rate constant of the linkers & $0.01  - 0.1 ~\mathrm{ sec ^{-1}}$  \cite{Alert2015-zq,Tsai2018-op,Fritzsche2014-nh} \\
$k_{scramblase}$ & Rate constant PS flipping through scramblase & $10  - 12$ ~1/(\textmu M s) \cite{Watanabe2018-gt} \\
$K_{m}$ & Rate constant of PS hydrolysis & $5.12 ~\mathrm{ sec ^{-1}}$  \cite{Watanabe2018-gt} \\
$k_{P4ATPase}$ & rate constant for P4ATPase activity & $1.5  - 2.0$ 1/(\textmu M s)  \cite{Watanabe2018-gt} \\
 &   & \\
\hline
\end{tabular}
\end{table}

\section*{Supplementary figures}
\subsection*{Effect of curvature-driven feedback on MP formation}
\begin{figure}[!!h]
    \centering
    \includegraphics[width=\textwidth]{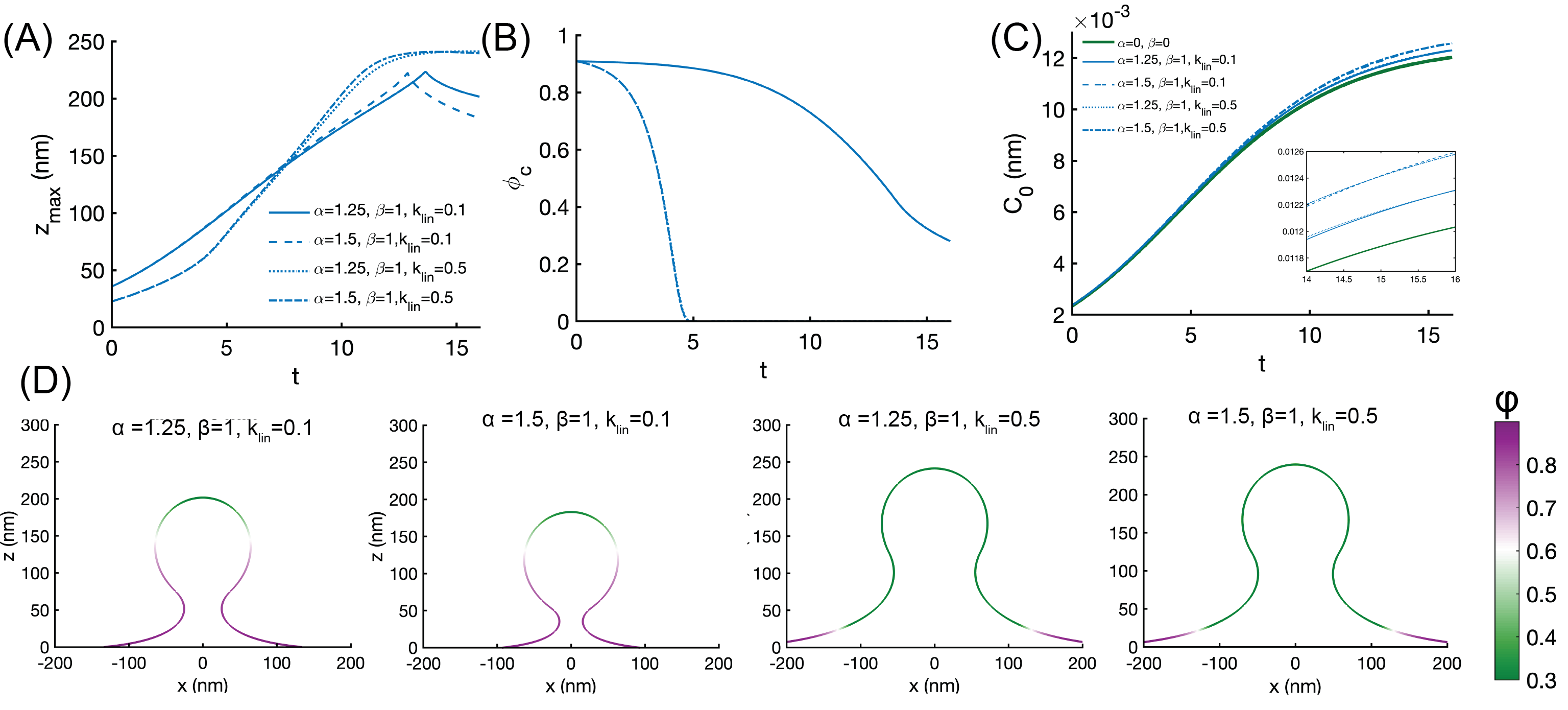}
    \caption{\textbf{Effect of curvature-driven feedback.}
    We used non-zero values of $\alpha$ and $\beta$ in \Cref{eq:PS_curvature_feedback}. We tested two different values of $\alpha$ for $beta=1$ because the exact values of these parameters are not known. (A) Change of maximum displacement of the membrane as a function of time (in min) for four different combinations $\alpha$ and $\beta$. (B) Change in $\phi$ at the center of the membrane as a function of time (min) for the combinations of $\alpha$ and $\beta$ shown in (A). (C) The curvature driven feedback serves to very slightly alter the kinetics of PS-induced spontaeous curvature. (D) Membrane shapes at final time for different values of $\alpha$, $\beta$, and $k_{lin}$ are shown.
  }
    \label{fig:curvature_feedback}
\end{figure}

We investigated how the strength of curvature driven feedback on PS kinetics could influence successful MP formation.
To do this, we took conditions that were already favorable for successful MP formation from \Cref{fig:PS_kinetics,fig:linker_stiffness}, which are WT PS kinetics and $\zeta=1$ and now tested how the strength of curvature dependent feedback ($\alpha$ and $\beta$) and linker stiffness play a role in successful MP formation (\Cref{fig:curvature_feedback}).
In these simulations, $\alpha$ was chosen to be greater than $\beta$ to ensure that scramblase activity was dominant and that PS-induced spontaneous curvature would be similar to the WT kinetics.
We observed that the maximum deformation at the center of the membrane depended on the value of linker stiffness and changes to the value of curvature driven feedback had a small effect \Cref{fig:curvature_feedback}A. 
The fraction of bound linkers depended on the value of linker stiffness (\Cref{fig:curvature_feedback}B).
The effective change in spontaneous curvature was not large in the presence of curvature driven feedback (\Cref{fig:curvature_feedback}C) and in all cases tested, we observed successful MP formation (\Cref{fig:curvature_feedback}D).

\subsection*{Effect of bending modulus and membrane tension}
\begin{figure}[!!h]
    \centering
    \includegraphics[width=\textwidth]{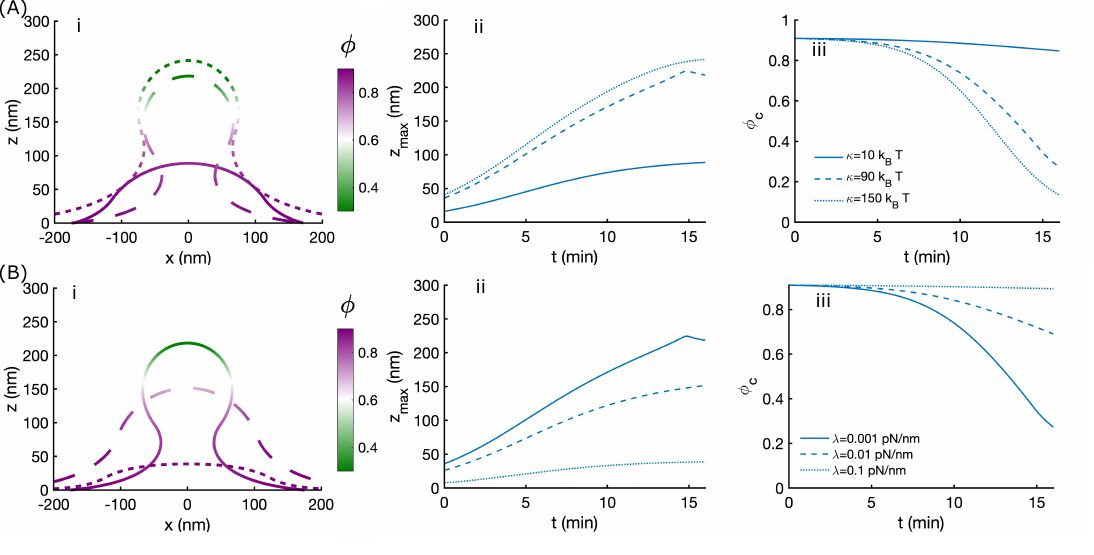}
    \caption{\textbf{Effect of bending modulus and membrane tension at $\zeta=0.1$ for WT PS kinetics.}
      (A) Effect of membrane bending rigidity on MP formation. Comparison of (i) equilibrium shapes, (ii) temporal evolution of membrane deformation ($z_{max}$), and (iii) temporal evolution of bound linker density at the center of the membrane for three different values of bending modulus  ($\kappa=10, 90, 150 k_BT$), with tension $\lambda = 0.001 pN/nm$ and $k_{lin}=0.1$ pN/nm. 
    (B)  Effect of membrane tension on MP formation. Comparison of (i) equilibrium shapes, (ii) temporal evolution of membrane deformation ($z_{max}$), and (iii) temporal evolution of linker density at the center of the membrane for three values of  membrane tension ($\lambda= 0.001, 0.01, 0.1 pN/nm$ ) with  $\kappa= 90 k_BT$ and $k_{lin}=0.1$ pN/nm.
  }
    \label{fig:membrane_properties}
\end{figure}

Next, we studied the role of membrane parameters like bending modulus $\kappa$ and the membrane tension $\lambda_0$ applied at the boundary. We see here the lower values of bending modulus and higher values of membrane tension are not favorable for forming microparticles (\Cref{fig:membrane_properties}(A) I, (B)i).
The dependence of membrane tension is monotonic; the lower the membrane tension, the easier it is to form microparticles. This follows the intuition of membrane deformation: higher tension makes the membrane stretched, and that resists bending of the membrane caused by spontaneous curvature of the pulling force.
However, the dependence of the bending modulus is not monotonic, as we see at the equilibrium intermediate value of bending rigidity $(90~ k_BT )$ from the closed bud. To understand this, we need to understand the role of the pulling force applied by the linkers and spontaneous curvature separately.
The role of pulling force is straightforward: bending modulus resists the protrusion of the tube; the lesser its magnitude, the easier it can protrude. 
Meanwhile, the dependence on spontaneous curvature is counter-intuitive. Higher bending rigidity helps in forming closed buds for a fixed spontaneous curvature. We explained this in great detail in \cite{Mahapatra2023-ez}.


\subsection*{Effect of fluctuation and noise}
\begin{figure}[!!h]
    \centering
    \includegraphics[width=\textwidth]{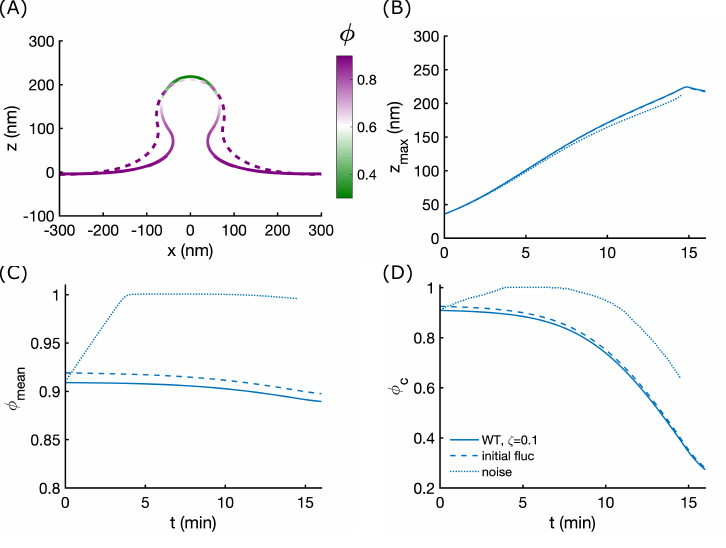}
    \caption{\textbf{Effect of noise and fluctuation for WT PS and $\zeta=0.1$.}
       Effect of membrane fluctuation and noise on MP formation. Comparison of (A) equilibrium shapes, (B) temporal evolution of membrane deformation ($z_{max}$), (C) Average density of the linkers', and (D) temporal evolution of bound linker density at the center of the membrane for WT PS kinetics and $\zeta=0.1$ ($\kappa=90k_BT$), with tension $\lambda = 0.001$ pN/nm and $k_{lin}=0.1$ pN/nm and  and three different situations, no noise and fluctuation, with initial fluctuation, and initial and spatio-temporal fluctuation of linkers' density.
  }
    \label{fig:noise}
\end{figure}

We also studied the stability of the shapes with respect to initial fluctuation and spatio-temporal noise in \Cref{fig:noise}. 
We noticed that the initial fluctuation in linker density does not affect the membrane deformation, equilibrium shapes, and temporal variation of $z_{max}$ follow the shapes without any initial fluctuation. 
There is a slight increase in linker density, but the nature of temporal dependence remains identical.
However, if we add spatial noise in linker density at each time, the average linker density shoots up, causing a deviation in membrane deformation temporally and leading to a different equilibrium shape.
This merits further exploration and the development of advanced stochastic solvers. 

\label{sec:model development}

\printbibliography




\end{document}